\documentclass[twocolumn,aps,prb]{revtex4}% Physical Review B

\usepackage{graphicx}% Include figure files

\begin{document}

\title{In-plane optical conductivity of La$_{2-x}$Sr$_x$CuO$_4$:
Reduced superconducting condensate and residual Drude-like response}
\author{S. Tajima$^{1)}$} 
\email{tajima@istec.or.jp}
\author{
Y. Fudamoto$^{1)}$\cite{address1}, T. Kakeshita$^{1,2)}$,
 B. Gorshunov$^{3)}$\cite{address2},  V. \v{Z}elezn\'{y}$^{1)}$\cite{address3}, 
K. M. Kojima$^{2)}$, M. Dressel$^{3)}$, and S. Uchida$^{2)}$ }

\affiliation{$^{1)}$Superconductivity Research Laboratory, ISTEC, 1-10-13 
Shinonome, Tokyo 135-0062, Japan\\
$^{2)}$Dept. of Physics, The University of Tokyo, Tokyo 113-8656, 
Japan\\
$^{3)}$I. Physikalisches Institut, Universit\H{a}t Stuttgart, 
D-70550 Stuttgart, Germany}

\date{\today}

\begin{abstract}
Temperature dependences of the optical spectra of La$_{2-x}$Sr$_x$CuO$_4$ with
 $x$=0.12 and 0.15 were carefully examined for a polarization parallel to the
 CuO$_2$-plane over a wide frequency range down to 8 cm$^{-1}$.  
Selection of well-characterized crystals enabled us to measure purely
 in-plane polarized spectra without any additional peak.  
The weight of superconducting (SC) condensate estimated from the missing
 area in $\sigma_1(\omega$) well agrees with the estimate from 
the slope of $\sigma_2(\omega$) vs 1/$\omega$ plot,
showing no evidence that the Ferrell-Glover-Tinkham sum-rule is violated
in the optical spectrum.
We demonstrate that the optically estimated SC condensate
is much smaller than the value obtained from the $\mu$SR measurement of 
magnetic penetration depth. 
We also find an anomalous increase of
 conductivity in sub-millimeter region towards $\omega$=0 below $T_c$, 
which suggests the microscopic inhomogeneity in the superconducting state.
Both observations are discussed in relation with the inhomogeneous electronic
 state that might be inherent to high-$T_c$ cuprates.

\end{abstract}

\pacs{74.25Gz, 74.72Dn, 71.27+a, 74.80.-g}

\maketitle 

\section{Introduction}
The superconducting (SC) condensate,
one of the essentail parameters for a superconductor,
has been a subject of discussion in high-$T_c$ superconducting cuprates(HTSC).
The recent progress in scanning tunneling spectroscopy has revealed a strange
 inhomogeneous distribution of the gap in Bi$_2$Sr$_2$CaCu$_2$O$_8$
 (BSCCO)\cite{davis1}.  
The effect of this real-space gap inhomogeneity on a superconducting (SC)
 condensate is unclear.  
On the other hand, the angle-resolved photoemission spectroscopy (ARPES) has
 demonstrated a distinct $k$-dependence of the gap at the Fermi surface in the
 pseudo-gap state or the stripe ordered state\cite{norman,shen}.
For the system with such a partially gapped Fermi surface, 
the source of SC condensate is not obvious.

As one of the direct methods to detect a SC gap and the weight of SC condensate,
 optical studies have been extensively performed for HTSC.  
A clear conductivity depression was observed below $\sim$500 cm$^{-1}$ in 
YBa$_2$Cu$_3$O$_y$ (YBCO)\cite{schlesinger,schuetzmann,basov4,wang} and 
BSCCO\cite{romero}.
Then, the weight of SC condensate ($\rho_s \sim \omega_{ps}^2)$ was
 estimated from the missing area ($A$) of the real part $\sigma_{1}(\omega)$ of
 conductivity spectrum or from the low-frequency behavior of its imaginary part
 $\sigma_2(\omega)$.

The estimate of SC condensate from optical spectra, however,
 has been subject to uncertainty. 
One of the reasons is that, unlike the case of conventional superconductors,
 there often remains some conductivity below the gap energy, 
forming a Drude-like increase towards $\omega$=0\cite{timusk}.
Since the missing area of $\sigma_1(\omega)$ is large in YBCO and BSCCO, 
the remaining conductivity effect is relatively small in estimation of SC
 condensate.  
By contrast, in the case of lower $T_c$-HTSC such as La$_{2-x}$Sr$_x$CuO$_4$
 (LSCO), 
it must give more serious effect.  
Nevertheless, this residual conductivity was ignored in most of the previous
 estimate of SC condensate except for a few works\cite{gorshunov,somal}.
The origin of this remaining conductivity is an important problem to be solved.

The other reason for the difficulty in determining SC condensate of LSCO is
 linked to the difficulty in obtaining a pure in-plane spectrum.  
Since the early stage of HTSC research, 
various different in-plane spectra have been reported for
 LSCO\cite{gorshunov,somal,gao,startseva,lucarelli} and
 La$_2$CuO$_{4+\delta}$\cite{quijada}.  
Some of the reported in-plane spectra show rich features in a far-infrared (FIR)
 region \cite{startseva,lucarelli,quijada}, 
which are ascribed either to the LO-phonons\cite{startseva} or to the excitation
 related to the polarons\cite{lucarelli}.  
According to the variety of spectra, 
different values of the SC condensate were derived, 
ranging from 250 to 430 nm\cite{gorshunov,somal,gao,startseva,lucarelli,quijada}
in terms of the London penetration depth 
$\lambda_L^{FIR}$(=1/2$\pi\omega_{ps}$), 
which do not correlate with the $T_c$ -value of the studied sample. 
Some of them are considerably different from the value determined by $\mu$SR
 measurements\cite{aeppli,uemura}.
  
The purpose of this work is two-fold.  
The first one is to clarify the origin of such a confusing situation in optical
 study of LSCO.  
We report a purely in-plane polarized spectrum of LSCO and 
demonstrate the effect of the $c$-axis component mixing, 
by comparing with the "dirty" spectra contaminated by the $c$-axis component.  
Possible reasons for the $c$-axis component mixing are discussed.  
The second and main purpose is to examine the electrodynamic response in the low
 energy region and to estimate the SC condensate,
based on the purely in-plane polarized optical spectra of
 La$_{2-x}$Sr$_x$CuO$_4$ (x=0.12 and 0.15).
The spectra were measured over a wide range of frequency 
8 - 30000 cm$^{-1}$, 
down to the sub-millimeter region.  
At $T \ll T_c$, 
we have found a depression of conductivity below 200 cm$^{-1}$ (70 cm$^{-1}$)
 for x=0.15 (x=0.12) but a steep increase of the conductivity below 
40cm $^{-1}$(10 cm$^{-1}$),
which leads to a significantly small missing area compared to the value 
estimated from the London length $\lambda_L$ determined by
 $\mu$SR measurement. 
We conclude that a discrepancy of the SC condensate between the optical and
$\mu$SR-estimations is commonly observed in many HTSC.
Possible scenarios for these anomalous features are discussed.

\section{Experiments}

Single crystals of La$_{2-x}$Sr$_x$CuO$_4$ with x=0.12 and 0.15 were grown 
by a traveling solvent-floating zone (TSFZ) method.
The superconducting transition temperature T$_c$ is 36K for x=0.15
and 30K for x=0.12.   
Fixing crystal axes by observing X-ray Laue patterns at the surface of as-grown
 crystal rods, 
we cut the crystals along the $c$-axis and took several pieces of samples with 
a thickness of about 2 mm. 
We call them "$ac$-face" sample hereafter. 
For comparison, we also cut some crystal pieces with the $c$-axis perpendicular to the measurement surfaces called "$ab$-face" sample.  
In both cases, $a$- and $b$-axes are not distinguished although there is a small
 difference between them.  
After a number of trial measurements of far-infrared spectra, 
we realized that even Laue-characterized crystals were sometimes multi-domains,
 showing the $c$-axis phonon admixture. 
 
The second characterization method to detect multi-domain is to observe the
 surface using a polarized optical microscope. 
Prior to this observation, 
the sample surfaces were polished with Al$_2$O$_3$ powder of different particle
 sizes at several steps. 
The particle diameter for the final polishing was 0.3 $\mu$m.  
It was found that some pieces of crystals consisted of multi-domains.  
We carefully selected the single domain crystals for measurement of purely
 in-plane polarized spectrum.  
A typical diameter of the sample disks is about 5 mm.  
A spot size of the incident light is about 3 mm, 
well smaller than the sample size. 

Optical reflectivity spectra were measured using a coherent-source 
spectrometer\cite{kozlov}
 in a millimeter and sub-millimeter wavelength region (8 - 33 cm$^{-1}$), 
a Fourier transformation type spectrometer for infrared region (30 - 
8000 cm$^{-1}$) and a grating type one for higher energy region above 
near-infrared (4000 - 30000 cm-1).  
Both a sample and a gold mirror were mounted on a copper plate in a He-gas flow
 cryostat for temperature control.   
Before each measurement of a sample spectrum, 
a spectrum of gold mirror was measured as a reference.  
Reflectance of polarized light was measured with the s-polarized geometry at the 
incident angle of about 5 degree.

In order to compare the SC condensate estimated from the other experimental 
methods, 
we measured the London penetration depth using muon spin rotation ($\mu$SR)
 technique. 
To improve the reliability of the missing area analysis and discussion, 
the $\mu$SR measurement has been carried out on the same single crsytal as used
 in the optical study.  
The analysis of the field distribution of a vortex lattice in the applied 
transverse field $H_t$=0.2 $Tesla$ has been performed using a London model 
\cite{fesenko} with a Lorentzian cut-off\cite{clem} of 4.5-6.5 nm as a non-local 
correction.  
Other possible contributions for the broadening of the field distribution, 
such as random pinning, 
are corrected by assuming an additional Gaussian relaxation with the relaxation 
rate 0.39-0.43 MHz.

\section{Results and Analysis}

\subsection{Effect of $c$-axis component admixture}

Figure 1 shows the far-infrared reflectivity spectra at room temperature for
 several samples.  
LSC\#1 is the almost perfectly in-plane polarized spectrum for the $ac$-face of
 La$_{1.85}$Sr$_{0.15}$CuO$_4$ crystal, while LSC\#2, 
the spectrum for another $ac$-face La$_{1.85}$Sr$_{0.15}$CuO$_4$ crystal, 
shows the bumps centered at around 300 cm$^{-1}$ and 520 cm$^{-1}$.  
LNSC\#1, measured without polarizer for the $ab$-face of 
La$_{1.45}$Nd$_{0.4}$Sr$_{0.15}$CuO$_4$, 
also shows a similar bump-like feature at the same position. 
Compared to the $c$-polarized spectrum ($E \parallel c$) in Fig.1, 
these bumps can be assigned to the two A$_{2u}$ mode phonons with the 
TO-frequencies of 240 cm$^{-1}$ and 500 cm$^{-1}$
that are the $c$-axis vibration modes of oxygens\cite{collins}.  
Namely, the bumps result from mixing of the $c$-polarized component into the 
in-plane spectrum.

%Fig.1
\begin{figure}%
\begin{center}
  \includegraphics[width=6cm]{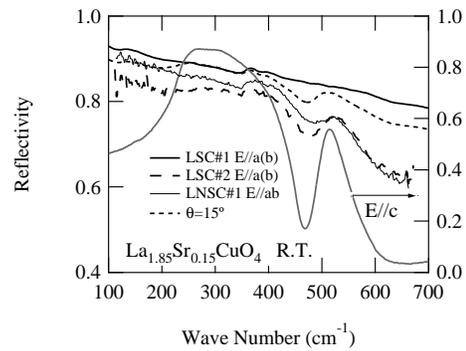}
  \end{center}
\caption{Comparison of the in-plane reflectivity spectra of 
three La(Nd)$_{1.85}$Sr$_{0.15}$CuO$_4$ crystals at 300K.
LSC\#1 and LSC\#2 were measured on the $ac$-faces,
while LNSC\#1 is the spectrum for the $ab$-face measured without polarizer.
The spectra of LSC\#2 and LNSC\#1 show a substantial contribution 
from the $c$-axis spectrum.  
The $c$-axis spectrum of LSCO is indicated by a dashed curve
as a reference.
We can reproduce a similar profile to LSC\#2 and LNSC\#1,
using the pure in-plane spectrum (LSC\#1) and the $c$-axis spectrum,
as described in the text.
The calculated reflectivity assuming a 15 degree rotation around the
$ac$-/$bc$-plane or equivalently the 6.7\%-admixture of the $c$-axis 
component is indicated by a dash-and-dot curve.}
\label{fig1}
\end{figure}
%Fig.1

We can reproduce a similar spectrum to LSC\#2 and LNSC\#1, 
using LSC\#1 as a pure in-plane reflectivity data ($R_{a/b}$) and the measured 
$c$-polarized reflectivity data ($R_c$), 
with $R_{\theta} = R_{a/b}$cos$^2(\theta)$ + $R_c$sin$^2(\theta)$.  
The dashed curve in Fig.1(a) represents the calculated reflectivity spectrum for 
$\theta$=15 degree deviation of the polarization direction from the pure 
$a$($b$)-axis or equivalently the 6.7\%-admixture of mis-oriented volume. 
 Corresponding to the rapid drop of reflectivity above 600 cm$^{-1}$, 
the dip at $\omega \sim$ 470 cm$^{-1}$ and the suppression below 250 cm$^{-1}$
 in the $c$-axis spectrum, 
the weak reflectivity suppressions giving the bump features in LSC\#2 and 
LNSC\#1 are well reproduced in the calculated spectrum.

The admixture of the $c$-axis spectral component into the in-plane spectrum 
seriously influences the conductivity and scattering rate spectra calculated 
from the reflectivity spectrum via Kramers-Kronig (KK) transformation.  
As shown in Fig.2(a),  the $c$-component admixture creates the conductivity dip 
between 300 and 1000 cm$^{-1}$ and the peak around 500 cm $^{-1}$.
These additional features were observed in both of conductivity and reflectivity
spectra for the previous studies of
LSCO\cite{startseva,lucarelli} and La$_2$CuO$_{4+\delta}$ \cite{quijada},
which clearly indicates the admixture of the $c$-axis component.  
Figure 2(b) demonstrates that the double peak in 
1/$\tau(\omega$) reported by Startseva {\it et al.}\cite{startseva} can also be 
considered as the same admixture effect.

%Fig.2
\begin{figure}
\begin{center}
  \includegraphics[width=6cm]{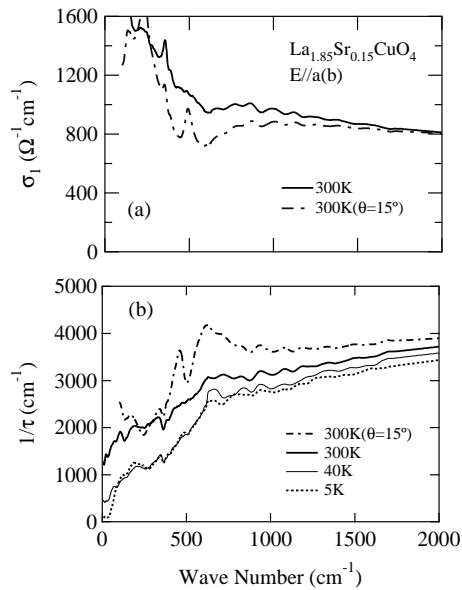}
  \end{center}
\caption{(a) Comparison of the in-plane conductivity $\sigma_1(\omega)$ of 
La$_{1.85}$Sr$_{0.15}$CuO$_4$ (LSC\#1) and the $c$-axis component mixed data 
for $\theta$=15 degree.
(b) Comparison of the scattering rate 1/$\tau(\omega)$ for LSC\#1 and
the $c$-axis component admixture case.
It is clear that the double peak is a result of the $c$-component admixture.
It is also demonstrated that the kink at $\sim$700 cm$^{-1}$ even at 300K and 
is independent of the superconducting response below 200 cm$^{-1}$.}
\label{fig2}
\end{figure}
%Fig.2

There are several possible reasons for the $c$-axis component mixing, 
such as (i)mis-cut of crystals, (ii)multi-domains in as-grown crystals, 
(iii)$p$-polarization admixture in non-polarized measurement.  
Mis-cut of as-grown crystal is the most primitive mistake that gives a 
mis-oriented surface for reflectivity measurement.  
This problem is peculiar to the LSCO crystals, 
because the cutting procedure is inevitable for extracting an appropriate size 
sample from an as-grown crystal rod. In the case of YBCO crystals grown by a 
flux method, natural surfaces can be used for optical measurement, 
and the BSCCO crystals grown by a TSFZ method are so easily cleaved that crystal 
cutting is not necessary to extract the $ab$-face samples.

The multi-domain problem is serious in the TSFZ crystals.  
It often appears in the early stage of crystal growth, namely, 
in the crystal portion close to a seed.  
It is hard to confirm that a sample is of single domain by a usual X-ray Laue 
pattern measurement, 
because a typical X-ray spot size is much smaller than the sample surface area 
of about 3x3 mm$^2$.  
This was the case for LSC\#2.  
In order to scan the whole area of sample surface, 
we need an X-ray topography measurement.  
An easier way to examine the existence of multi-domain feature is to observe a 
polished sample surface using a polarized optical microscope.  
In the worst case, even the sample selected by this method shows some admixture 
of different direction component in its far-infrared spectrum.  
The mis-oriented domain behind the surface within a penetration depth 
($\sim$ 1 $\mu$m) of far-infrared light is a possible origin.  
The dielectric response of the light penetrating through multi-domains must be 
complicated, 
giving an unusual structure in far-infrared spectra.  
The obtained reflectivity cannot merely be interpreted as a weighted summation 
of the in-plane component $R_{a/b}$ and the $c$-axis component $R_c$ in this 
case.

The final problem related to the $p$-polarization must be considered when we 
measure the in-plane spectrum on the $ab$-face using non-polarized light.  
In order to gain a signal, we sometimes did not use a polarizer for the 
$ab$-face measurement because the light intensity is reduced to about a half 
when going through a polarizer.  
However, this geometry measurement always gave a weak bump-like feature that 
roughly corresponds to the $c$-axis phonon peak or high-frequency reflectivity 
suppression.  
An example is shown in Fig.1(a) as LNSC\#1 for the $ab$-face of 
(La,Nd,Sr)$_2$CuO$_4$ crystal which was well characterized by an X-ray 
diffraction.  
A similar problem can be seen in the set of data by Lucarelli {\it et al.}
\cite{lucarelli}
About a half of the seven spectra were measured on the $ab$-face 
without a polarizer,
and they show a clear peak corresponding to the $c$-axis phonon\cite{tajima}.
The most plausible origin of this $c$-axis component mixing is the 
$p$-polarization component in the non-polarized light, 
as van der Marel {\it et al.} pointed out \cite{marel}.  
Although the introduced $c$-component is very small if the surface is perfectly 
$ab$-oriented, 
the effect possibly becomes evident in extremely anisotropic materials like 
HTSC.

All the problems due to the $c$-axis component mixing become serious only in 
strongly anisotropic systems, namely, when the reflectivity values are 
remarkably different in the $ab$- and $c$-polarized spectra. 
This is well demonstrated in the optical spectra of intentionally mis-alighed 
films of (La,Ce)$_2$CuO$_4$\cite{pimenov}, 
where only 1 degree mis-orientation results in a clear feature of the $c$-axis 
phonons in the spectra. 
This remarkable effect is due to the completely insulating spectral profile for 
$E \parallel c$.  
In fact, less anisotropic materials such as well-oxygenated YBCO have never 
shown the $c$-axis component in the spectrum measured with the in-plane 
polarized light.  
Therefore, the measurement of in-plane spectrum for LSCO and probably 
(Nd,Ce)$_2$CuO$_4$ requires particularly high quality samples without any 
multi-domain feature, 
high accuracy in cutting angle and careful polarization measurements.  
These requirements are quite sever, 
compared to that for other measurement techniques such as neutron scattering.

\subsection{In-plane optical spectra}

%Fig.3
\begin{figure}
\begin{center}
    \includegraphics[width=6cm]{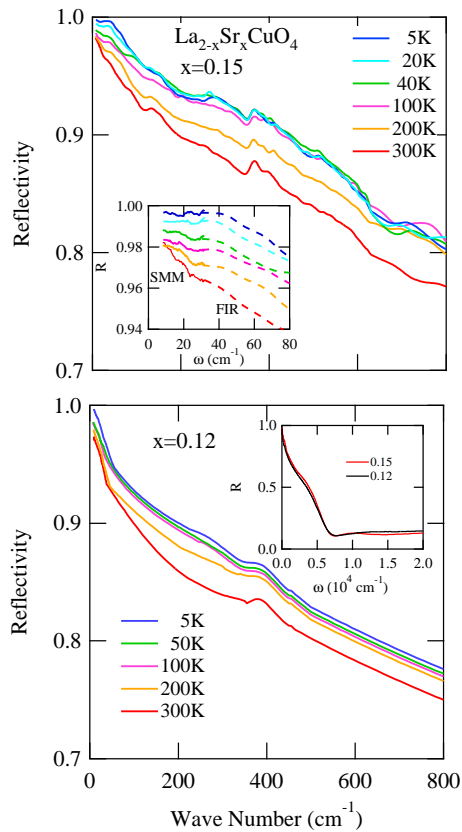}
  \end{center}
\caption{Temperature dependence of in-plane reflectivity spectra of 
La$_{2-x}$Sr$_{x}$CuO$_4$ with x=0.15 and x=0.12 with $E \parallel a$ or $b$. 
The inset of the upper panel shows the spectra for x=0.15 in
 sub-millimeter (SMM) and far-infrared (FIR) regions separately.
The inset of the lower panel shows the room temperature spectra 
in a wider range of frequency.
}
\label{fig3}
\end{figure}
%Fig.3

Hereafter we focus on the results of the sample for LSC\#1 which we believe the 
purely in-plane polarized spectrum, 
and for comparison on the results of the x=0.12 sample with 
$ac$-face which was also well characterized.  
Figure 3 shows the in-plane reflectivity spectra for x=0.15 (a) and 0.12 (b) at 
various temperatures.  
Phonon peaks are almost screened by free carriers except for the 
small structure at $\sim$370 cm$^{-1}$ ascribed to the in-plane oxygen bending 
mode phonon.
No admixture of the $c$-axis component is observed in both sample spectra.  
The FIR reflectance is smoothly connected to the sub-millimeter reflectance as 
shown in the inset of Fig.3(a).  
When the temperature is lowered below $T_c$, 
the reflectance for x=0.15 exhibits a rapid increase below 150 cm$^{-1}$ 
and a plateau below 40 cm$^{-1}$, 
which could be originating from the condensation of low frequency spectral 
weight transferred to $\delta(\omega$) at $\omega$=0.  
A similar feature is observed in the in-plane SC spectra of YBCO and BSCCO, 
but at higher frequency of $\sim$500 cm$^{-1}$, 
probably owing to the larger energy scale of SC gaps.

%Fig.4
\begin{figure}
\begin{center}
   \includegraphics[width=6cm]{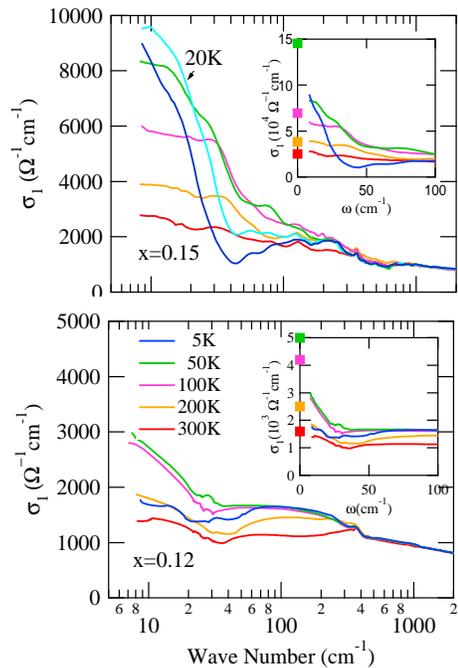}
  \end{center}
\caption{Temperature dependence of in-plane conductivity spectra of 
La$_{2-x}$Sr$_{x}$CuO$_4$ with x=0.15 and x=0.12 with $E \parallel a$ or $b$. 
Inset figures show the low-$\omega$ spectra and the dc conductivity.
The optical data can be smoothly extrapolated to the dc values.
}
\label{fig4}
\end{figure}
%Fig.4

The reflectance is transformed to the complex conductivity $\sigma(\omega$) 
by utilizing KK relations. 
Figure 4 shows the real part conductivity $\sigma_1(\omega$) that
 can be smoothly extrapolated to a 
$dc$ value at each temperature above $T_c$ as shown in the inset, 
justifying the validity of our KK analysis.  
The low-$\omega$ $\sigma_1(\omega$) for x=0.15 increases with reducing the temperature, 
forming a Drude-like peak at $\omega$=0, as is seen in Fig.4(a).  
When the temperature decreases below $T_c$, 
$\sigma_1$ starts to be suppressed below 250 cm$^{-1}$, 
corresponding to the SC gap opening. 
This SC response is, however, 
followed by an unusual Drude-like increase below 40 cm$^{-1}$ even at 5K$\ll T_c$.  
A similar Drude-like response is also seen in the low-$\omega$ upturn
of the spectrum for x=0.12 (Fig.4(b)).  
The Drude-like peak width is narrower in the x=0.12 sample than that for x=0.15.
A huge amount of residual conductivity at $T \ll T_c$ has also been reported
in a direct measurement of conductivity of La$_{1.84}$Sr$_{0.16}$CuO$_4$ film 
in the submillimeter wavelength region\cite{gorshunov}.
Moreover, this feature seems to be common in many HTSC such as YBCO and 
BSCCO\cite{timusk},
which should affect the estimation of SC condensate.

To compare with the previous results of LSCO\cite{startseva}, YBCO\cite{basov}
 and BSCCO\cite{puchkov}, 
we derive an $\omega$-dependent scattering rate 
1/$\tau(\omega)=(\omega_p^2/4\pi)Re[1/\sigma(\omega$)] for x=0.15 as plotted in 
Fig.2(b). 
Here $\omega_p$ is obtained from the integral
$\int_0^{\omega_0}\sigma_1(\omega)d\omega$=$\omega_p^2$/8.
We take $\omega_0$=10000 cm$^{-1}$ so that the integral range includes the 
reflectance plasma edge, 
but is well below the charge transfer excitation energy in La$_2$CuO$_4$\cite{uchida}. 
The spectrum of 1/$\tau(\omega$) clearly shows a kink at 650 cm$^{-1}$ at all 
temperatures.  
The kink in 1/$\tau(\omega$) is commonly observed at nearly the same frequency 
for all HTSC studied so far\cite{puchkov}.  
This has been ascribed to the opening of a pseudogap above $T_c$ and/or the 
coupling of carriers with a magnetic collective mode observed as a resonance 
peak in neutron scattering\cite{neutron}.  
Recently another candidate of source for this kink has been proposed by the 
study of angle-resolved photoemission spectroscopy (ARPES)\cite{lanzara}.  
It was demonstrated that the kink in energy-dispersion curve of ARPES is 
observed at nearly the same energy in all HSTC, 
irrespective of presence of the magnetic resonance peak. 
A phonon mode is, then, proposed as the most likely origin of the kink in 
energy-dispersion curve of ARPES and 1/$\tau(\omega$).

The present result confirms that the kink in 1/$\tau(\omega$) exists at 700-800 
cm$^{-1}$ commonly in HTSC.  
A difference of our result for LSCO from those of YBCO and BSCCO is that 
the kink is observed even at room temperature in the optimally doped sample with 
x=0.15.  
The pseudogap temperature at this composition is expected to be 70-80K from the 
$T$-dependences of in-plane resistivity and magnetization\cite{oda}.  
It means that the kink structure is not related to the pseudogap opening.  
Another important difference is the SC-gap related suppression which
is observed at $\omega\sim$100 cm$^{-1}$ that is far below the kink frequency.  
This is in sharp contrast to the continuous development from a pseudogap to a SC 
gap observed in the other HTSC such as YBCO and BSCCO\cite{basov}.  
The present observation indicates that the kink feature is irrelevant
to a SC gap.

\section{Discussions}

%Fig.5
\begin{figure}
\begin{center}
   \includegraphics[width=6cm]{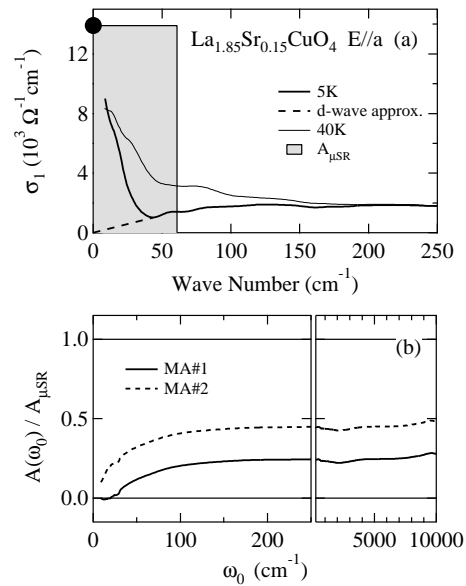}
  \end{center}
\vspace{0.5cm}
\caption{(a) Optical conductivity of La$_{1.85}$Sr$_{0.15}$CuO$_4$
($T_c$=36K) with $E \parallel a$ or $b$ at $T$=40K and 5K.
The missing area $A_{\mu SR}$ expected from $\lambda_L$=280 nm obtained in the
$\mu$SR study is shown by a hatched area.
The solid circle indicates the dc conductivity at 40K. 
(b) The ratio of the missing area obtained from the present optical study 
$A(\omega_0)$ to $A_{\mu SR}$.
MA\#1 represents the ratio for $A(\omega_0)$ calculated with the lower
limit of integral of 8 cm$^{-1}$.
MA\#2 is for the data of $A(\omega_0)$ calculated by the dashed curve extrapolation in Fig.5(a).
}
\label{fig5}
\end{figure}
%Fig.5

\subsection{SC condensate}
The observed SC response is most naturally interpreted as the signature of 
formation of SC condensate associated with a SC 
gap opening.  
When a delta-function peak develops at $\omega$=0 in $\sigma_1(\omega$) at the 
expense of its suppression in the low-$\omega$ region, 
$\sigma_2(\omega$) behaves as proportional to $\omega_{ps}^2/\omega$.  
There are two ways to estimate the SC condensate $\omega_{ps}^2$ 
from the optical spectrum.  
One is to calculate the missing area $A(=\omega_{ps}^2$/8) in 
$\sigma_1(\omega$), 
and the other is to estimate $\omega_{ps}$ from the slope of $\sigma_2(\omega$) 
vs 1/$\omega$.  
In the case of HTSC, 
it is not a trivial issue that these two methods give an identical value of 
$\omega_{ps}$,
since a possibility of kinetic energy contribution has been suggested by both theory\cite{hirsch} and experiment\cite{molegraaf,santander1}
We compare the two optically obtained values to check the Ferrell-Glover-Tinkham
(FGT) sum rule, adn also compare them 
with those estimated from the London 
penetration depths that 
were determined by $\mu$SR measurement.  

Below, we focus on the condensate weight at the lowest temprerature ($\sim$5K)
 well below $T_c$.
The conductivity missing area for x=0.15 is defined by the integral 
\begin{equation}
A(\omega_0)=
\int_{0+}^{\omega_0}[\sigma_1(\omega,40K)-\sigma_1(\omega,5K)]d\omega,
\end{equation} 
where we set $\omega_0$= 10000 cm$^{-1}$.  
We first put the lower limit of the integral at 8 cm$^{-1}$ which is the limit 
of our optical measurement (labeled as MA\#1).  
In Fig.5(a), the missing area $A_{\mu SR}$=1/8(2$\pi\lambda_L^{\mu SR})^2$
 expected from $\lambda_L^{\mu SR}$=280 nm 
obtained by $\mu$SR study\cite{kadono} is shown by the hatched area.  
It is obvious that $A_{\mu SR}$ is much larger than the missing area in the 
optical conductivity spectrum between 40 and 5K.  
The ratio of the optical missing area $A(\omega_0$) and the $\mu$SR value
 $A_{\mu SR}$ is plotted in Fig.5(b) as a function of $\omega_0$.  
$A(\omega_0)/A_{\mu SR}$ increases and then saturates to 0.25$\pm$0.05, 
showing no appreciable change up to 10000 cm$^{-1}$.  
The penetration depth corresponding to this value of $A(\omega_0$) is 530 nm.
Even if we eliminate the contribution of the Drude-like component at 5K and 
replace it by a straight line (shown by a dashed line in Fig.5(a)) 
which mimics a d-wave SC behavior, 
$A(\omega_0)/A_{\mu SR}$ amounts only to 0.5.  
This is plotted as a dashed curve labeled as MA\#2 in Fig.5(b), 
giving an upper limit of the missing area (or the lower limit of 
$\lambda_L^{FIR}$=400 nm) in our experiment.

The anomalously small missing area is also observed for x=0.12.  
Compared to the SC condensate estimated from the 
$\mu$SR $\lambda_L^{\mu SR}$=310 nm, 
the optical conductivity missing area $A(\omega_0$) is substantially small with 
the ratio $A(\omega_0$=10000 cm$^{-1}$)/$A_{\mu SR}\sim$0.1.  
The optically determined penetration depth turns out to be $\sim 1\mu$m
 in this case.

Strictly speaking, the missing area which is connected with the SC condensate
 should be calculated from the difference 
between the normal and the superconducting conductivity at $T$=5K.
In the above estimation of $A_0(\omega_0)$,
we assume that  $\int\sigma_1(\omega,40K)d\omega$=$\int\sigma_1^n(\omega,5K)d\omega$,
where $\sigma_1^n(\omega,5K)$ is the conductivity in the case that the system keeps the normal state down to $T$=5K.
In a conventional metal(superconductor),
this assumption holds well, 
because the low-$T$ resistivity is almost constant 
(the residual resistivity regime).
In the case of LSCO with x=0.12 and 0.15,
the 'normal state' resistivity realized by the application of intense
magnetic fields increases weakly
 with lowering $T$ below $T_c$\cite{ando},
suggestive of a 'normal' state rather similar to the conventional one.
As regards the observed $T$-dependence of the normal-state spectrum,
Fig.4 (and its inset) indicates that the frequency range where $\sigma_1$
shows an appreciable $T$-variation shrinks with decreasing $T$.
For the spectrum at $T$=50K,
a difference from that at $T$=100K is seen restricted below 30 cm$^{-1}$.
The trend suggests that the $T$-dependence of $\sigma_1^n$ arises 
predominantly from the Drude term.
From this, and also from the result of dc resisitivity measured under
intense fields,
we guess that, if the normal state persists below $T_c$,
the Drude term would not change significantly from that at $T_c$,
and hence the difference between $\sigma_1$(40K) and $\sigma_1^n$(5K)
would be small.

Another problem in estimation using eq.(1) is the possible contribution
from the higher frequency spectrum.
Recently the spectral weight reduction in the high energy region 
above 10000 cm$^{-1}$ has been observed in BSCCO by two independent methods
and the change in kinetic energy at superconducting transition has been 
reported\cite{molegraaf,santander1}.
We have also found small $T$-dependence of the present spectra in the visible 
frequency region,
which gives additional weight to $A(\omega_0$=10000cm$^{-1}$).
However, the integral up to $\omega_0$=30000 cm$^{-1}$ increases $A(\omega_0$)
only by 20\% at most.
The kinetic energy change, if it exists, 
cannot explain the small missing area in the present results.

%Fig6
\begin{figure}
\begin{center}
  \includegraphics[width=7.5cm]{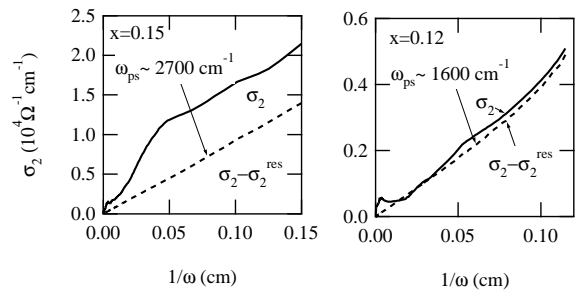}
  \end{center}
\caption{Imaginary part of condutivity $\sigma_2(\omega)$ of 
La$_{1-x}$Sr$_{x}$CuO$_4$ for x=0.12 and 0.15 at low frequencies.
$E \parallel a(b)$ and $T$=5K.
$\sigma_2^{res}$ is calculated from the residual $\sigma_1(\omega)$ at 5K
using KK-relation.
When $\sigma_2^{res}$ is substracted,
$\sigma_2$ well scales with 1/$\omega$,
which is the signature of superconducting carrier response.
}
\label{fig6}
\end{figure}
%Fig6

The second method to estimate the SC condensate is to use $\sigma_2$.
In Fig.6, $\sigma_2$ is plotted as a function of 1/$\omega$.
It is substantially deviated from the linear-relationship,
because large amount of remaining conductivity $\sigma_1(\omega)$
 below 50 cm$^{-1}$ strongly influencs $\sigma_2(\omega$).  
In order to remove the contribution of this residual conductivity, 
we adopt the method proposed by Dordevic {\it et al.}\cite{dordevic}.  
First, we assume $\sigma_1(\omega$) at 5K as the conductivity of unpaired 
carriers $\sigma_1^{res}(\omega)$ which do not condense at $\omega$=0. 
Next, we calculate $\sigma_2^{res}(\omega$) corresponding to 
$\sigma_1^{res}(\omega$) using a KK-relation.  
Then, the $\sigma_2^{s}(\omega$) which should represent the ''true'' superfluid 
response is obtained as 
$\sigma_2^{s}(\omega$) = $\sigma_2(\omega)-\sigma_2^{res}(\omega$).

The obtained $\sigma_2^{s}(\omega$) is proportional to 
1/$\omega$ both for x=0.12 and 0.15, 
indicating a SC response in this frequency range.  
The SC condensates giving $\lambda_L^{FIR}$=990 nm for x=0.12 and 590 nm for x=0.15 
are estimated from the slopes, respectively. 
These are in agreement with the values (1 $\mu$m for x=0.12 and 530 nm 
for x=0.15) obtained from the missing area in $\sigma_1(\omega$).  
It means that this method for estimation of SC condensate is self-consistent 
within the optical spectra, 
and thus it is concluded that the FGT sum rule almost holds within an infrared 
frequency range.
The present result does not completely exclude a possibility of small 
contribution from the high-$\omega$ 
region, namely, the kinetic energy contribution to the condensate weight,
as was observed in BSCCO\cite{molegraaf,santander1},
but it is beyond the accuracy of our measurement.

The previously reported values of $\lambda_L^{FIR}$ for the optimally doped LSCO 
range from 250 nm to 430 nm, 
depending on the sample and the research group but not on the $T_c$-value.  
The value estimated from the spectrum showing the $c$-axis
 component\cite{startseva,quijada} is not reliable, 
because the $c$-axis admixture modifies the conductivity spectrum as 
demonstrated in Fig.1.  
In fact, the estimated $\lambda_L^{FIR}$s' from their data are shorter 
than ours.

%table1
\begin{table*}
\caption{Comparison of penetration depths estimated from FIR and
$\mu$SR measurements.\\
(The FIR data of YBa$_2$Cu$_3$O$_y$ are for $E\parallel a$, 
while the $\mu$SR data are the average within the $ab$-plane.)}
\begin{ruledtabular}
\begin{tabular}{lccccc}
 &\multicolumn{2}{c}{La$_{2-x}$Sr$_x$CuO$_4$} & 
\multicolumn{2}{c}{YBa$_2$Cu$_3$O$_y$} & Bi$_2$Sr$_2$CaCu$_2$O$_z$ \\ 
 & x=0.15 & x=0.12 & y$\approx$6.9 & y$\approx$6.6 & $T_c\approx$70K \\
\hline
$\lambda_L^{FIR}$(nm) & 400-530 & 630\footnote{from Ref.39}-990 & 
160\footnote{from Ref.7}
 & $\sim$300\footnote{from Ref.42} & 680\footnote{from Ref.43} \\
$\lambda_L^{\mu SR}$(nm) & 280\footnote{from Ref.34} & 310\footnote{from Ref.17}
 & 112\footnote{from Ref.41} & 170\footnotemark[7] & 190\footnotemark[6] \\
\end{tabular}
\end{ruledtabular}
\end{table*}
%table1

Another point to judge the data quality is whether the low-$\omega$ Drude-like 
component is observed and taken into account or not.  
If the remaining conductivity in the SC state is ignored, the missing area must 
be overestimated.  
The relatively larger missing area (or equivalently the shorter 
$\lambda_L^{FIR}$=250nm), 
reported in the paper by Gao\cite{gao} may partly result from this 
over-estimation of the missing area, 
because the 5K-$\sigma_1(\omega$) was set to zero below 70 cm$^{-1}$.  
The measurement by Gorshunov {\it et al.} \cite{gorshunov} on the 
La$_{1.84}$Sr$_{0.16}$CuO$_4$ film ($T_c$=39.5K)\cite{film}, 
which covered the frequency range down to 5 cm$^{-1}$, 
clearly detected a narrow Drude-like band with the large weight 
$\omega_p\sim$7800 cm$^{-1}$ at 5K.  
Subtracting the Drude-like spectral contribution by fitting, 
they obtained a long penetration depth $\lambda_L^{FIR}$ =400 nm.  
A little longer penetration depth $\lambda_L^{FIR}$=430 nm was reported by Somal 
{\it et al.}\cite{somal} for a single crystal.  
These values are in fairly good agreement with our estimate of the shorter limit 
of $\lambda_L^{FIR}$.
Therefore, we conclude
 that the spectra successfully measured without contamination of the $c$-axis 
component give a similar value of $\lambda_L$ (400-500 nm) that is longer 
than the estimate (250-310 nm) from the contaminated spectra.

The microwave (MW) conductivity measurement gives a similar large value 
(=$\lambda_L^{MW}$=400$\pm$100 nm for x=0.15) to our FIR result, 
although the error bar is quite large in this technique\cite{shibauchi}.
By contrast, the $\mu$SR penetration depth
\cite{aeppli,uemura} is clearly shorter than 
the FIR and microwave values.
For x=0.12, our estimate from the FIR data is $\lambda_L^{FIR}$=990 nm,
the recent FIR data for x=0.125 by Dumm {\it et al.}\cite{dumm} shows 
$\lambda_L^{FIR}$=630 nm, 
and the microwave data for x=0.12 is $\lambda_L^{MW}$=500$\pm$200 nm
\cite{shibauchi}.
Note that $\lambda_L^{FIR}$ as well as $T_c$ is very sensitive to the Sr-content
x around x=0.12 owing to the so-called 1/8-anomaly ($T_c$ is most suppressed
when x is tuned to 0.115).
Even if taking into account this situation,
 we can conclude that the values of $\lambda_L^{FIR}$
 are much longer than the $\mu$SR estimate 
$\lambda_L^{\mu SR}$=310 nm.

A similar discrepancy between the FIR- and the $\mu$SR-penetration depths
is also seen in the case of YBCO and BSCCO particularly 
in the underdoped regime.
In the optimally doped YBCO, the average value of $a$- and $b$-polarized
data ($\lambda_L^{FIR}$=160 nm for $E \parallel a$ and 117 nm for $E 
\parallel b$)\cite{basov4,wang} is larger than the $\mu$SR value 
[$\lambda_L^{\mu SR}(ab)$=112 nm\cite{sonier}].
Comparing the conductivity spectra of optimally and under-doped 
YBCO\cite{rotter,basov},
we find that the missing area of $\sigma_1$ in the underdoped YBCO with 
$T_c$=56-59K is about 20\% of that for the optimally doped one with 
$T_c$=93K.
This suggests 
the penetration depth of about 300-400 nm for the underdoped YBCO.
The recent estimation of $\lambda_L^{FIR}$ by subtracting the low-$\omega$
residual conductivity is about 250 nm for YBa$_2$Cu$_3$O$_{6.6}$\cite{homes}.
This is much longer than the $\mu$SR value ($\lambda_L^{\mu SR}$ = 170 nm)\cite{sonier}.
A longer FIR penetration depth is also reported for underdoped BSCCO
with $T_c$=70K\cite{santander2}.
Compared with the $\mu$SR penetration depth ($\sim$190 nm)
 for BSCCO with $T_c$=75K\cite{uemura},
the reported value of $\lambda_L^{FIR}$ (=680 nm) for the $T_c$=70K 
sample\cite{santander2} is extremely long.
We summarize the comparison of the FIR- and $\mu$SR-penetration depths
for various HTSC in Table I.
As is clearly seen in the table, 
the discrepancy between the FIR and $\mu$SR data is quite robust for
HTSC.

One may consider
the possibility that a FIR spectrum is not
correctly measured by some reason,
for example, by the [110] surface giometry problem arising from the $d$-wave 
symmetric gap,
 as recently pointed out by Tu {\it et al.}\cite{tu}.
It is, however, unlikely because a small estimate of SC condensate was
obtained by various measurements with various surface giometries.

It is also a general trend that the discrepancy becomes larger as one goes to more underdoped regime.
This is suggestive of electronic inhomogeneity
as a possible origin of the discrepancy.
The SC order parameter may not be uniform in
real sapce and vanish in some parts, according to the recent observation
of STM on BSCCO\cite{davis1}. 
In LSCO, the stripe fluctuation may introduce electronic inhomogeneity.
As we discuss below, the presence of non-superconducting area appears to be
correlated with the residual conductivity in the SC state,
which is responsible partly for the reduced missing area in $\sigma_1(\omega$).

A distinct difference from the $\mu$SR results was recognaized in the 
impurity-substituted YBCO\cite{wang}, YBa$_2$Cu$_4$O$_8$
and many other disordered HTSC\cite{basov2}.
Disorder, impurity or defect, decreases the FIR SC condensate 
($\lambda_L^{FIR})^{-2}$ 
at much steeper rate than that expected from the linear 
$T_c$-$(\lambda_L^{\mu SR})^{-2}$ relationship observed by $\mu$SR.
For example,
0.4\% Zn-substitution for optimally doped YBCO suppresses $T_c$
from 93K to 85K and duplicates $\lambda_L^{FIR}$ ($\sim$340 nm 
for $E\parallel a$),
while $\lambda_L^{\mu SR}$ is expected to increase only 20\% at most
by the same Zn-substitution\cite{nachumi}.
The discrepancy between the FIR and $\mu$SR data
seems to be universal in the disordered or inhomogeneous cuprates,
including the underdoped cuprates for which STM suggests an inherent
spatial modulation of the SC order parameter over a lenght scales of 
a few nm.
It may follow that the FIR measurement is more sensitive to microscopic
inhomogeneity than $\mu$SR,
and that the ordinary relationship between $\rho_s$ and $\lambda_L$ 
($\rho_s \sim \lambda_L^{-2}$) no more holds for such microscopically
inhomogeneous SC-state.

\subsection{Low-$\omega$ residual conductivity at $T \ll T_c$}

Finally we discuss the residual conductivity in the SC state.
The Drude-like up-turn of conductivity towards $\omega$=0 is observed
almost in all HTSC even at $T$ well below $T_c$\cite{timusk}.
In the present study, the low-$\omega$ Drude-like spectral weigh at $T \ll T_c$ 
is smaller for x=0.12 than for x=0.15,
the Drude-like peak width being narrower in the former.
This is consistent with the general trend that the spectral weight of the 
residual conductivity increases with doping\cite{corson}.

The low-$\omega$ conductivity remaining at $T \ll T_c$ is an indication of the 
presence of non-superconducting region.
So far known are three types of inhomogeneous state in HTSC:
(i) Presence of non-SC region within the SC area,
either metallic as speculated for overdoped cuprates by FIR\cite{schuetzmann2}
and $\mu$SR measurements\cite{niedermeyer,uemura2}
or pseudogapped as observed by STM for underdoped BSCCO\cite{davis1},
(ii) Alternating array of anti-ferromagnetic (AF) and SC 
stripes\cite{tranquada},
(iii) Local suppresion of the SC order around impurities such as Zn, 
as was directly observed by STM\cite{pan}.

The impurity-induced inhomogeneity [the case (iii)] has been investigated 
by many experimental techniques.
In the optical spectra, 
Zn-substitution creates a huge residual conductivity,
which dramatically depresses the missing area\cite{wang}.
A similar observation was reported for the Zn-doped 
YBa$_2$Cu$_4$O$_8$\cite{basov2} and in the irradiated YBCO\cite{basov3}, 
giving the evidence for the sensitiveness of the FIR probe to microscopic 
inhomogeneity.

The case (ii), the spin and charge stripe order can be considered as an 
''ordered'' inhomogeneous state,
 which has been detected by neutron scattering experiment as a static order 
in (La,Nd,Sr)$_2$CuO$_4$\cite{tranquada}, 
and probably as a dynamical form in LSCO\cite{yamada}, 
where the Sr-content x=0.12($\sim$1/8) is expected to be closer to the static 
order. 
In this case, the charge stripes on which SC oreder might develop are separated
by the AF spin domains, and would form a periodic Josephson coupled array.
A weak spatial modulation of the Josephson coupling strength between the 
stripes, due to e.g. stripe meandering,
is a possible source of the low-$\omega$ conductivity peak,
as was observed in the $c$-axis optical response\cite{marel2,kakeshita,zelezny}.
When the Josephson coupling strength is increased and strongly modulated
spatially by decreasing an average spacing between stripes and/or by enhanced
stripe fluctuation,
the weight of residual conductivity would increase and be distributed
over a fairly wide frequency range.

\section{Conclusion}

The in-plane polarized spectra of LSCO with x=0.12 and 0.15 were carefully 
measured.  
The peak at $\sim$500 cm$^{-1}$ that was observed in the previous reports and/or 
in some of our samples was assigned to the $c$-axis phonon mode ($A_{2u}$). 
Various possible sources for this admixture of $c$-axis spectral component were 
examined.  
Inaccurate angle of crystal cutting and/or multi-domains in the TSFZ-crystals 
possibly introduces some amount of the dielectric response for $E\parallel c$.  
Measurement on $ab$-face with non-polarized light or with $p$-polarization 
geometry also results in the $c$-axis component mixing.  
In any case, the problem originates from the large anisotropy in the electronic 
system of LSCO where only a small amount of $c$-axis component makes a serious 
effect on the in-plane spectrum, and therefore, 
this is the problem peculiar to a strongly anisotropic material like LSCO.

The obtained pure $a$/$b$-axis spectrum showed a clear superconducting response, 
the suppression of $\sigma_1(\omega$) and 1/$\omega$-behavior of 
$\sigma_2(\omega$).  
The estimation both from $\sigma_1$ and $\sigma_2$ give a consistent value of
SC condensate,
which indicates that the kinetic energy contribution is not appreciable.
We find that the SC condesate is much smaller than that determined by $\mu$SR.  
This discrepancy is possibly caused by microscopic inhomogeneity in
the electronic state of supercnducting CuO$_2$-planes, 
probably related to the stripe fluctuation in the case of LSCO.
It is guessed that FIR-measurement is a probe sensitive to disorder
which suppresses the SC order parameters over a length scale of nano-meter.
The inhomogeneous electronic state also seems to manifest in the spectrum
 as a residual Drude-like 
response at very low frequencies below 50 cm$^{-1}$ in the SC state.

\section*{Acknowledgement}
This work was supported by the New Energy and Industrail Technology Development 
Organization (NEDO) as Collaborative Research and Development of Fundamental 
Technologies for Superconductivity Applications, and by a COE Grant and a 
Grant-in-Aid For Scientific Research on Priority Area from the Ministry of 
Education, Japan.


\begin{references}
\bibitem{film} The film is subject to strain from the substrate which makes 
$T_c$ higher than that for unstrained single crsytal presumably due to 
suppression of stripe fluctuation.
This would make $\omega_{ps}(\lambda_L^{FIR})$ of the film higher(shorter).
\end{references}
\end{document}